\documentclass[12pt,a4paper]{article}
\usepackage{a4wide}
\usepackage{makeidx}
\usepackage{latexsym}
\usepackage{epsf}
\usepackage{amssymb}
\makeindex

\begin{document}
\def\be{\begin{equation}}
\def\ee{\end{equation}}
\def\bea{\begin{eqnarray}}
\def\eea{\end{eqnarray}}

\def\pd{\partial}
\def\a{\alpha}
\def\b{\beta}
\def\g{\gamma}
\def\d{\delta}
\def\m{\mu}
\def\n{\nu}
\def\t{\tau} 
\def\l{\lambda}
\def\s{\sigma}
\def\e{\epsilon}
\def\scri{\mathcal{J}}
\def\cM{\mathcal{M}}
\def\tcM{\tilde{\mathcal{M}}}
\def\RR{\mathbb{R}}
\def\CC{\mathbb{C}}

\hyphenation{re-pa-ra-me-tri-za-tion}
\hyphenation{trans-for-ma-tions}


\begin{flushright}
IFT-UAM/CSIC-01-21\\
hep-th/0106195\\
\end{flushright}

\vspace{1cm}

\begin{center}

{\bf\Large Topics in String Tachyon Dynamics}
\footnote{Lectures at the IV SIGRAV School and 2001 School of Algebraic 
Geometry and Physics}
\vspace{.5cm}

{\bf C\'esar G\'omez}
\footnote{E-mail: {\tt cesar.gomez@uam.es  }}
{\bf and Pedro Resco }
\footnote{E-mail: {\tt juanpedro.resco@uam.es }} \\
\vspace{.3cm}

\vskip 0.4cm

\
Instituto de F\'{\i}sica Te\'orica, C-XVI,
  Universidad Aut\'onoma de Madrid \\
 { E-28049-Madrid, Spain}\footnote{Unidad de Investigaci\'on Asociada
  al Centro de F\'{\i}sica Miguel Catal\'an (C.S.I.C.)}

\vskip 1cm


{\bf Abstract}

We review some aspects of string tachyon dynamics with special emphasis on 
effective actions and K-theory interpretation.

\end{center}


\begin{quote}

\end{quote}


\newpage

\setcounter{page}{1}
\setcounter{footnote}{1}
\tableofcontents
\newpage
\section{Introduction}

In the last few years a new understanding of the dynamical role 
of tachyons in string theory has started to emerge (\cite{1} to \cite{17}). For 
the simplest open bosonic string there exist already a lot of evidence 
on tachyon condensation \cite{1,6,7,8,11,12}. 
The tachyon vacuum expectation value 
characterizing this condensate exactly cancel the open string one 
loop contribution to the cosmological constant, what we now understand as the 
$D_{25}$ filling brane tension. The vacuum defined by this condensate 
is naturally identified with the closed string vacua. Precise computations 
of the tachyon potential supporting this picture has been carried out both 
in open string field theory \cite{7,8,9,10,12,13,14,15} and in 
background independent open 
string field theory \cite{18, 19, 20, 21, 22, 23}. At this level of understanding 
two main problems remain open. First of all we have the problem of 
the closed tachyon that survives as an unstability of the closed string 
vacua defined by the open string tachyon condensate. Secondly we lack a 
precise understanding of the dynamical mechanism by which the $U(1)$ 
gauge open degrees of freedom are decoupled from the closed string spectrum. 

Concerning the problem of the closed string tachyon, the $\sigma$-model beta 
functions \cite{24, 25} indicate that closed tachyon condensation creates 
a contribution to the cosmological constant of the same type generated by 
working with non critical dimensions. The well known result about the $c=1$ 
barrier in the context of linear dilaton backgrounds \cite{26} could 
indicate a sort of unstability that reduces drastically the space time 
dimensions until reaching the safe $D=2$.

With respect to the problem of the fate of $U(1)$ gauge degrees of 
freedom after open tachyon condensation - a short of confinement of 
open degrees of freedom into closed spectrum - there are two formal hints. 
One is the suggestion of a trivial nilpotent BRST charge of type $ac_{0}$, 
for $c$ the ghost field, around the background defined by the tachyon 
condensate \cite{27}. The other hint comes from observing that the open 
string effective Born-Infeld lagrangian, is multiplied by a factor $e^{-T}$ 
with $T=\infty$ defining the open tachyon condensate \cite{28, 29, 30, 23}.

In the context of more healthy superstrings without tachyons, the phenomena 
of tachyon condensation shed some new light on the solitonic interpretation 
of the $D$-branes. We have two main examples corresponding to pairs 
$D_{p}-D_{\bar{p}}$ brane-antibrane which will support an open tachyon on 
the world volume spectrum and the case of configurations of unstable non 
BPS $D$-branes. In both cases tachyon condensation will allow us to 
interpret stable BPS $D$-branes as topologically stable extended objects, 
solitons, of the auxiliary gauge theory defined on the world volume of 
the original configuration of unstable $D$-branes.

The mechanism of decay into a closed string vacua by tachyon 
condensation can be used to define a new algebraic structure to characterize 
$D$-brane stability and $D$-brane charges, namely K-theory 
(\cite{31} to \cite{35}). 
The main ingredient in order to go to K-theory is the use of stability 
equivalence with respect to creation-annihilation of branes. In type IIB   
$D_{p}$-branes of space codimension $2k$ are related to $K(B^{2k},S^{2k-1})$ 
and for type IIA $D_{p}$-branes of space codimension $2k+1$ are related 
to $K^{-1}(B^{2k+1},S^{2k})$. The characterization of $K(X,Y)$ in 
terms of triplets \cite{36} $(E,F,\a)$ with $E,F$ vector bundles on $X$ and 
$\a$ an isomorphism $\a :E\mid_{Y}\rightarrow F\mid_{Y}$ makes specially 
clear the mathematical meaning of the open tachyon field as defining the 
isomorphism $\a$. A similar construction in terms of pairs $(E,\a )$ with $\a$ 
an automorphism of $E$ can be carried out for the definition of the higher 
$K^{-1}$-group \cite{32}.

Finally we would like to point out to some striking similarities between 
the topological characterization of stable $D_{p}$-branes in type IIA 
string and gauge fixing singularities for unitary gauges \cite{37}  of the type of 
'tHooft's abelian 
projection \cite{38}. Can we learn something of dynamical relevance from this 
analogy?. After the discovery of asymptotic freedom, the Holy Grial of high 
energy physics is the solution of the confinement problem. The abelian 
projection gauge was originally suggested in \cite{38} as a first step 
towards a quantitative approach to confinement i.e. to the computation of 
the magnetic monopole condensate. The analogy between stable $D_{p}$-branes 
$(p\leq 6)$ in type IIA and the magnetic monopoles associated with 
the abelian projection gauge singularities seems to indicate, as the 
stringy analog of confinement, the decay of the gauge vacua associated with a 
configuration of unstable $D_{9}$-filling branes into a closed string 
vacua populated of stable $D_{p}$-branes. Another interesting lesson we learn 
from the analogy is that as it is the case with magnetic monopoles in the 
abelian projection, that should be considered as physical degrees of 
freedom independently of what is the phase, confinement, Higgs or Coulomb 
of the underlying gauge theory, the same should be true concerning 
$D_{p}$-branes in type IIA, independently of the concrete form of the 
open tachyon potential. What is relevant to characterize the ``confinement'' 
closed string phase is the ``dualization'' of the original open gauge string 
degrees of freedom into RR closed string fields whose sources are stable 
$D_{p}$-branes. Finally and from a different point of view another hint 
suggested by this analogy is the potential relevance of the higher K-group 
$K^{-1}$ to describe gauge fixing singularities in ordinary gauge theories. 
Maybe the answer to the natural question why the higher K-group 
$K^{-1}$ is pointing out to some hidden ``M-theoretical'' meaning of 
the gauge $\theta$-parameter.

The present review is not intended to be complete in any sense. Simply 
covers the material presented by one of us (C.G.) during the $4^{th}$ SIGRAV  
School on Contemporary Relativity and Gravitational Physics and 2001 School 
on Algebraic Geometry and Physics. Como May 2001
\footnote{Some parts of these lectures were also presented in:
II Workshop on Non-commutative Geometry, String Theory, and 
Particle Physics. Rabat May 2001 and in the 
Workshop New Interfaces between Geometry and Physics. 
Miraflores June 2001. 
}.

\section{Why tachyons?}

In quantizing string theory in flat Minkowski space-time there are two
 constants that should be fixed by consistency, namely the normal
 ordering constant appearing in the mass formula: 
\be\label{string_mass}
M^{2}=\frac{4}{\a'}(N-a)
\ee
and the dimension D of the space-time. These two constants 
determines the Virasoro anomaly
\be
[L_{m},L_{n}]=(m-n)L_{m+n}+A(D,a,m)\d_{m+n}
\ee
with
\be
A(D,a,m)=\frac{D}{12}(m^3-m)+\frac{1}{6}(m-13m^3)+2am
\ee
and 
\be
L_{m}=L_{m}^{(matter)}+L_{m}^{(ghosts)}-a\d_{m}
\ee
Impossing $A(D,a,m)=0$ implies the standard constraints on the bosonic 
string, namely $D=26$ and $a=1$.

The first consequence of the non vanishing normal ordering constant $a$ is
 that the $(mass)^2$ of the ground state $(N=0)$ is 
negative i.e. it is a tachyon.
In spite of that there is a good consequence of this normal ordering
 value, namely, the existence at the first level of a massless vector
 boson in the open case and the massless graviton in closed case.

A priori the only consistency requirement we should imposse is absence of
 negative norm ghost states in the physical Hilbert space. This allow
 us to relax the condition on D and $a$ to $D\leq 26$ and $a\leq 1$.

Although in these conditions the open string theory is perfectly healthy
 at tree level we will find unitarity problems for higher order corrections,
 more precisely singularity cuts for one loop non planar diagrams. 
In the closed string case the problems at one loop will show up 
as lack of modular invariance.
 Thus we will reduce ourselves to critical
 dimension $D=26$ and $a=1$.

One important place where the normal ordering constant appears
 in string theory is in the definition of the BRST operator:     
\be\label{brst_charge}
Q=\sum_{m}(L_{m}c_{-m}-\frac{1}{2}\sum_{n}(m-n)c_{-m}c_{-n}b_{m+n})
\ee
with $b,c$ the usual ghost system for the bosonic string. The charge $Q$
 can be written in a more compact way as:
\be
Q=\sum_{m}(L_{m}^{(matter)}+\frac{1}{2}L_{m}^{(ghosts)}-a\d_{m})c_{-m}
\ee
Notice that the contribution of the normal ordering constant to $Q$ is simply 
$ac_{0}$. This quantity defines by itself a BRST charge -since it is trivially 
nilpotent $c_{0}^2=0$- with trivial cohomology\footnote{This is the BRST 
operator recently suggested in \cite{27} to describe the 
cohomology around the open tachyon condensate.}

In standard quantum fiel theory a tachyon is not such an unfamiliar object.
 A good example is for instance the Higgs field if we perturb 
around the wrong vacua
$\langle \phi \rangle =0$.
In this sense the presence of a tachyon usually means that we are 
perturbing around an unstable vacua. In a physically sensible situation
 we expect the system roll down to some stable vacua where automatically
 the tachyon will disappears. In the bosonic string it is not clear at all 
if this is the case since we still lack a powerful tool to study
the string theory off shell. The only real procedure to address this 
issue is of course string field theory.

In superstring theories with space-time supersymmetry i.e. type I, type II 
or heterotic, the tachyons are proyected out by impossing GSO. However 
even in these cases open string tachyons can appear if we consider non-BPS 
Dirichlet D-branes. In those cases the open tachyon is associated 
with unstabilities of these non-BPS D-branes.

\section{Tachyons in AdS: The $c=1$ barrier}

A simple way to see the unstabilities induced by tachyonic fields with
 negative $(mass)^2$, is to compute their contribution to the energy in 
flat Minkowski space-time. Generically the energy is defined by
\be
E=\int d^{n-1}x dr \sqrt{g}[g^{\m\n}\pd_{\m}\phi^{*}\pd_{\n}\phi+
          m^{2}\phi^{*}\phi]
\ee
where $n$ is the space-time dimension. The condition of finite energy 
requires an exponential falloff $\phi \sim e^{-\lambda r}$ 
at infinity with $\lambda > 0$. The energy of a field fluctuation with
 this falloff at infinity goes like $E \sim (\lambda^2 + m^2)$.
Thus if $m^2<0$ this energy can becomes negative for small enough 
$\lambda$ , which means unstability. This is not necessarily the case
if we consider curved space time.

For $AdS_{n}$ the metric can be written as :
\be 
ds^{2}=e^{2ky}dx^{2}_{n-1}+dy^2
\ee
with the curvature radius being:
\be
R=\frac{1}{k}
\ee
For simplicity let us consider fluctuations of the field depending only
 of the $y$ coordinate. The condition of finite energy requieres now
 an exponential falloff  $\phi \sim e^{-\lambda y}$ for $y\rightarrow \infty$
 with
\be
\lambda > \frac{k(n-1)}{2}
\ee
As before the contribution to the energy will go as
  $E \sim (\lambda^2  + m^2)$ and therefore we get positive energy for
tachyon fields with $m^2=-a$ if
\be
a\leq \frac{(n-1)^2}{4R^2}
\ee
This bound on the tachyon mass in $AdS_{n}$ is known as Breitenlohner-
Freedmann bound \cite{39}.

In the case of string theory the contribution to the energy of
  closed string tachyons goes like:
\be\label{umi}
E= \int d^{d-1}x dr \sqrt{g} e^{-2\Phi}[g^{\m\n}\pd_{\m}T\pd_{\n}T+
          m^{2}T^{2}]
\ee
with $m^2=-\frac{4}{\a'}$. The field $\Phi$ in (\ref{umi}) is the 
dilaton field.
We will be interested in working in flat Minkowski space-time of dimendion n. 
The dilaton $\sigma$-model beta function equation 
\be
\frac{n-26}{6\a'}+ (\nabla \Phi)^2 -\frac{1}{2}(\nabla ^2 \Phi)=0
\ee
implies a lineal dilaton behavior of type:
\be
\Phi = y\sqrt{\frac{n-26}{6\a'}}
\ee
for some arbitrary coordinate $y$. Let us now consider tachyon fluctuations 
on this background depending only on coordinate $y$. Using 
the same argument that with $AdS_{n}$ we get 
the bound on the tachyon mass $m^2=-a$ :
\be 
a \leq \frac{n-26}{6\a'}
\ee
Thus in order to saturate this bound for the closed string tachyon
 $a=\frac{4}{\a'}$ we need $n=2$. This is the famous $c=1$ barrier,
 namely only for space time dimension equal two or smaller the closed
 string tachyon is not inducing any unstability.

Notice that from the point of view of the tachyon mass bound, linear 
dilaton for dimension $n$ behaves as $AdS_{n}$ with curvature radius given 
by\footnote{For solutions to the bosonic beta function interpolating between AdS and linear dilaton see references \cite{37a, 37b}}:
\be
R^2= \frac{3(n-1)^{2}\a'}{2(n-26)}
\ee

\section{Tachyon $\sigma$-model beta functions}

The partition function for the bosonic string in a closed tachyon background
is given by:
\be\label{sigmaaction}
Z(T)=\int Dx e^{\frac{-1}{2\pi\a'}\int d^{2}\sigma 
\sqrt{h}(h^{\a\b}\pd_{\a}x^{\m}\pd_{\b}x^{\n}\eta_{\m\n}+T(x))}
\ee

The first thing we notice is that the tachyon term $\int\sqrt{h}T(x)$
is clearly non invariant with respect to Weyl rescalings of the
 world-sheet metric. The strategy we will follow would be to fix 
$h_{\a\b}=e^{2\phi}\eta_{\a\b}$ in (\ref{sigmaaction}) and to imposse invariance with respect 
to changes of $\phi$ for the quantum corrected $\sigma$ -model. We will use a 
background field $x_{0}^{\m}$ with $x^{\m}=x_{0}^{\m}+\xi^{\m}$ and such that
$\pd_{\m}T(x_{0})=0$. In these conditions we get at one loop in the
 $\sigma$-model:

\bea\label{effaction}
\frac{1}{2\pi\a'}\int d^{2}\sigma 
e^{2\phi}(T(x_{0})+\frac{\a'}{2}\pd_{\m}\pd_{\n}T(x_{0})<\xi^{\m}\xi^{\n}>+...)
\\
= \frac{1}{2\pi\a'}\int d^{2}\sigma 
e^{2\phi}(T(x_{0})+
\frac{\a'}{2}\pd_{\m}\pd_{\n}T(x_{0})\eta^{\m\n}\log\Lambda+...)
\eea

where by $<\xi^{\m}\xi^{\n}>$ we indicate the one loop of quantum fluctuations 
(see figure 1) and where $\Lambda$ is the ultraviolet cutoff for the
 one loop integration.

\begin{figure}
\begin{center}
\leavevmode
\epsfxsize=4cm
\epsffile{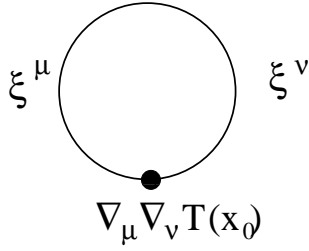}
\caption{\it One loop contribution to the tachyon beta function.}
\label{fig1}
\end{center}
\end{figure}

Next we need to relate the Weyl factor $\phi$ with the cutoff $\Lambda$.
A dilatation of the world-sheet metric induces a change $\Lambda \rightarrow
 \lambda \Lambda$ and $e^{\phi} \rightarrow \lambda e^{\phi}$,
thus we can identify $e^{\phi}$ with $\Lambda$. Doing this we get 
from (\ref{effaction}):
\be\label{expansion}
\Lambda^{2}[T(x_{0})+
\frac{\a'}{2}\pd_{\m}\pd_{\n}T(x_{0})\eta^{\m\n}\log\Lambda]
\ee
Expanding (\ref{expansion}) in powers of $\log\Lambda$ we get, at first order
 in $\log\Lambda$, that independence of Weyl rescalings requires
\be\label{tachyonequation}
\b_{T}\equiv 2T(x_{0})+
\frac{\a'}{2}\pd_{\m}\pd_{\n}T(x_{0})\eta^{\m\n}=0
\ee
which is the definition of the closed string tachyon beta function.

Repeating exactly the same steps for the open string tachyon we get
 instead of (\ref{tachyonequation}):

\be\label{tachyonequationopen}
\b_{T}^{o}\equiv T(x_{0})+
\a'\pd_{\m}\pd_{\n}T(x_{0})\eta^{\m\n}=0
\ee

If we iterpret (\ref{tachyonequation}) and (\ref{tachyonequationopen}) as equations of motion they correspond 
to tachyonic space time fields of $(mass)^2$ respectively
$-\frac{4}{\a'}$ and  $-\frac{1}{\a'}$.

What we learn from this simple exercise is that the tachyonic nature
of background T introduced in (\ref{sigmaaction}) is tied to 
the simple fact that 
$\int_{\Sigma}\sqrt{h}T$ is not Weyl invariant. Notice that although 
the usual dilaton term $\int_{\Sigma}\sqrt{h}\Phi R^{(2)}$ is not Weyl 
invariant it depends on $\phi$ only through $(\pd\phi)^2$ terms.

\section{Open strings and cosmological constant: The Fischler-Susskind mechanism}
\subsection{Fischler-Susskind mechanism: Closed string case}

Let us start considering one loop divergences in the critical $D=26$
 closed bosonic string. For simplicity we will reduce ourselves to 
amplitudes with M external tachyons. Divergences for this amplitude
 will arise in the limit where all the M external tachyon insertions
 coalesce (see figure 2).

\begin{figure}
\begin{center}
\leavevmode
\epsfxsize=6cm
\epsffile{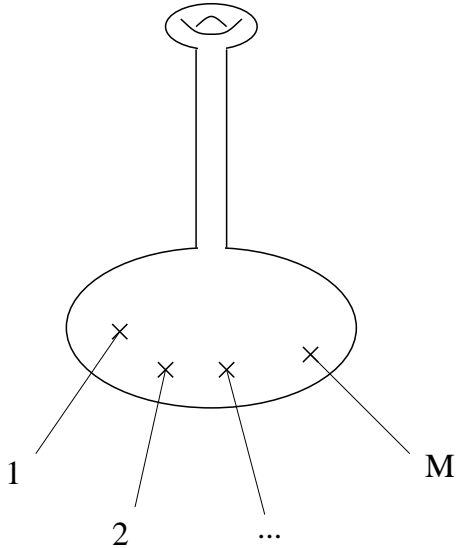}
\caption{\it Relevant topology to describe the limit where the insertion 
points coalesce.}
\label{fig2}
\end{center}
\end{figure}

The amplitude is given by:
\be\label{closedamplitude}
A(1,2,...,M)= \int \frac{d^2\tau}{(Im\tau)^2}C(\tau)F(\tau)
\ee
where
\be
C(\tau)= (\frac{Im\tau}{2})^{-12}e^{4\pi Im\tau}|f(e^{2i\pi \tau})|^{-48}
\ee
and
\be\label{vertexexpresion}
F(\tau)= \kappa^{M} Im\tau \int \prod^{M-1}d^{2}\n_{r}\prod_{r<s}
(\chi_{rs})^{\frac{k_{r}k_{s}}{2}}
\ee
Expresion (\ref{closedamplitude}) is invariant under $SL(2,Z)$ modular transformations
\be
\tau \rightarrow \frac{a\tau +b}{c\tau +d}
\qquad    ad-bc=1
\ee
Integration in (\ref{closedamplitude}) is reduced to the fundamental 
domain $F$.
Using the conformal Killing vector on the torus we have fixed the position 
$\n_{M}$ of one external tachyon. It is convenient to define the 
new variables:
\be
\varepsilon\eta_{r}\equiv \n_{r}-\n_{M}\qquad r=1,...,M-2
\ee
\be
\eta_{M-1}\equiv \n_{M-1}-\n_{M}=\varepsilon e^{i\phi}
\ee
with $\varepsilon$ and $\phi$ real variables. 
The jacobian of the transformation is:
\be
\prod^{M-1}d^{2}\n_{r}= i\varepsilon^{2M-3}d\varepsilon d\phi 
\prod^{M-2}d^{2}\eta_{r}
\ee
In the limit where $\n_{rs}=\n_{r}-\n_{s} \sim 0$ the Green function
 $\chi_{rs}$ in (\ref{vertexexpresion}) behaves like:
\be
\chi_{rs} \sim 2\pi |\n_{rs}|
\ee
Expanding in this limit the integrand in (\ref{vertexexpresion}) 
in powers of $\varepsilon$
 the leading divergence is:
\be\label{necktachyon}
\kappa^{M}\int_{0}^{1}\frac{d\varepsilon}{\varepsilon^{3}}d\phi \prod^{M-2}
d^{2}\eta_{r}\prod_{1\leq r \leq s \leq M-1}
|\eta_{r}-\eta_{s}|^{\frac{k_{r}k_{s}}{2}}\int  \frac{d^2\tau}{(Im\tau)}C(\tau)
\ee
where we have used the on shell condition for the closed tachyon
\be
\sum_{1\leq r \leq s \leq M-1}k_{r}k_{s}= -4M
\ee
The amplitude (\ref{necktachyon}) correspond 
to the propagation of a closed tachyon
 along the neck. The next subleading term in the expansion goes like
 $\frac{1}{\varepsilon}$ and correspond to the propagation along the 
neck of a massless dilaton. Thus divergent contribution to the amplitude 
can be written like:
\be\label{neckdilaton}
\int_{0}^{1}\frac{d\varepsilon}{\varepsilon}A_{0}(k=0,1...M)\kappa J
\ee  
where $A_{0}$ is the genus zero amplitude for M external tachyons and 
one dilaton at zero momentum and where $\kappa J$ is proportional to the 
genus one dilaton tadpole(see figure 2):
\be
\kappa J =\kappa \int_{F}  \frac{d^2\tau}{(Im\tau)^2}C(\tau)
\ee

\begin{figure}
\begin{center}
\leavevmode
\epsfxsize=6cm
\epsffile{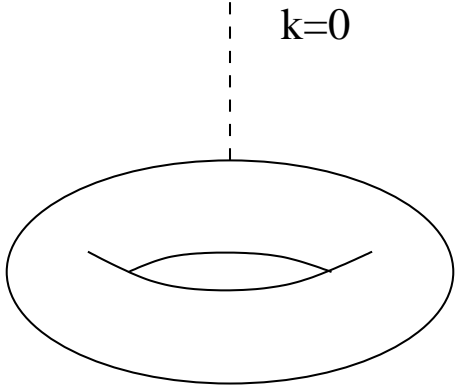}
\caption{\it Dilaton tadpole graph.}
\label{fig3}
\end{center}
\end{figure}

The original idea of Fischler-Susskind mechanism \cite{40} consist in 
absorbing the genus one divergence (\ref{neckdilaton}) into a 
renormalization of
 the world sheet $\sigma$-model lagrangian, namely: 
\be\label{renorm}
\eta_{\m\n}\pd x^{\m}\pd x^{\n}\rightarrow \eta_{\m\n}[1+
\kappa^{2}J\int_{0}^{1}\frac{d\varepsilon}{\varepsilon}]\pd x^{\m}\pd x^{\n}
\ee
The factor $\kappa^{2}$ in (\ref{renorm}) appears because we want to use this 
counterterm on the sphere to cancel a genus one divergence.Recall that
 generic genus one amplitudes goes like  $\kappa^{M}$ while genus
 zero amplitudes goes like $\kappa^{M-2}$. 

Obiously the renormalized lagrangian defined in (\ref{renorm}) explicetely
 breaks conformal invariance. Introducing a cutoff $\Lambda$ in the
 $\varepsilon$-integration the corresponding $\sigma$-model beta function is:
\be\label{betacorrection}
\b_{\m\n}^{(1)}=\kappa^{2}J\eta_{\m\n}\sim \frac{\d L_{R}}{\d\log \Lambda}  
\ee
for $ L_{R}$ the renormalized lagrangian defined in (\ref{renorm}). In principle 
we can generalize (\ref{betacorrection}) to curved space time just replacing $\eta_{\m\n}$ 
by $G_{\m\n}$. Once we do that we can compensate the   
 $\sigma$-model beta function coming from $\sigma$-model one loop effects:
\be
(\log \Lambda)R_{\m\n}\pd x^{\m}\pd x^{\n}
\ee
with the genus one contribution, by impossing:
\be\label{gequation}
R_{\m\n}=\kappa^{2}JG_{\m\n}
\ee
In summary the main message of Fischler-Susskind mechanism is that 
$\sigma$-model divergences can be compensated by string loop
 divergences. We have shown that this is th case case at least at genus one. Including 
the dilaton field and using the well known relation  
\be
\kappa =e^{\Phi} 
\ee
we will get instead of (\ref{gequation}):
\be
R_{\m\n}-2\nabla_{\m}\nabla_{\n}\Phi=e^{2\Phi}JG_{\m\n}
\ee

\subsection{Open string contribution to the cosmological
            constant: The filling brane}

This time we will consider the open string one loop amplitude for M external 
on shell open tachyons(see figure 4)

\begin{figure}
\begin{center}
\leavevmode
\epsfxsize=6cm
\epsffile{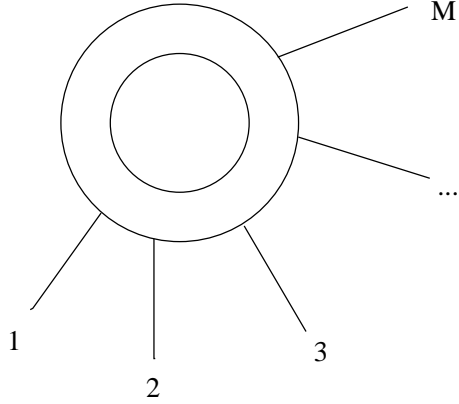}
\caption{\it One loop open string amplitude.}
\label{fig4}
\end{center}
\end{figure}

In the planar case this amplitude in given by :
\be
A(1,2,...,M)= g^{M}\int_{0}^{1}\prod^{M-1}\theta(\n_{r+1}-\n_{r})d\n_{r}
\int_{0}^{1}\frac{dq}{q} q^{-2}[f(q^2)]^{-24}\prod_{r<s}
[\Psi_{rs}]^{k_{r}k_{s}}
\ee
The divergences of this amplitude appear in the $q\rightarrow 0$ limit 
corresponding to shrinking to zero size the hole of the annulus. The 
structure of divergences can be read from the annulus vacuum to vacuum 
amplitude:
\be\label{annulus}
Z_{0}^{(1)}= \int_{0}^{1}\frac{dq}{q} q^{-2}[f(q^2)]^{-24}=
\int_{0}^{1}\frac{dq}{q^{3}}[1+(26-2)q^{2}+...]
\ee
Extending Fischler-Susskind to (\ref{annulus}) is equivalent to reproduce the coefficient 
of the divergences in terms of expectation values of certain operators 
on the disc \cite{41}. The divergence $26 \int_{0}^{1}\frac{dq}{q}$ is easely 
reproduced by: 
\be\label{opendiv}
\int_{0}^{1}\frac{dq}{q}\frac{e^{\Phi}}{\a'}
<\eta_{\m\n}\pd x^{\m}\pd x^{\n}>_{disc}
\ee
where we have included the dilaton factor requiered for matching 
the one loop and disc amplitudes. The divergence  
$\int_{0}^{1}\frac{dq}{q^{3}}$ correspond to:
\be
\int_{0}^{1}\frac{dq}{q^{3}}e^{\Phi}
<1_{d}>_{disc}
\ee
The logaritmic divergence  $-2 \int_{0}^{1}\frac{dq}{q}$ comes from the 
contribution of ghosts to the annulus partition function. The correct 
way to reproduce this divergence is in terms of the ghost dilaton vertex 
operator $D^{(ghost)}(k=0)$ as

\be\label{ghost}
\int_{0}^{1}\frac{dq}{q}e^{\Phi}
<D^{(ghost)}(k=0)>_{disc}
\ee
In fact the representation (\ref{ghost}) 
of the divergence $-2 \int_{0}^{1}\frac{dq}{q}$
 is a direct consequence of the dilaton theorem \cite{42}:
\be
<\int D_{ghost}(z,\bar{z})\Phi (p_{1})\dots \Phi(p_{n})>_{\Sigma} \sim
2g-2+n<\Phi (p_{1})\dots \Phi(p_{n})>_{\Sigma}
\ee
with
\be
D_{ghost}(z,\bar{z})=\frac{1}{2}c\pd^{2}c-\bar{c}\bar{\pd}^{2}\bar{c}
\ee
Let us concentrate on (\ref{opendiv}), the 
Fischler-Susskind counterterm needed 
to cancel this divergence induces a contribution to the $\b_{\m\n}$ 
$\sigma$-model beta function proportional to 
\be
\frac{e^{\Phi}}{\a'}\eta_{\m\n}
\ee
In order to reproduce this term we need to add to the closed string 
effective lagrangian the open string cosmological constant term:
\be\label{kappa}
\frac{1}{\kappa^{2}}\int d^{26}x \frac{e^{-\Phi}}{\a'}\sqrt{g}
\ee
The reader can easely recognize in (\ref{kappa}) the first 
term in the expansion 
of the $D-25$ filling brane Born-Infeld lagrangian: 
\be
S_{BI}= T_{25}e^{-\Phi}\int d^{26}x \sqrt{g+b+F}=
T_{25}e^{-\Phi}\int d^{26}x \sqrt{g}+...
\ee
with $T_{25}$ the filling brane tension given by 
$\sim \frac{1}{\a' \kappa^{2}}$.

Thus we learn that the $D-25$ filling brane tension simply represents 
the open string contribution to the cosmological constant.

Before finishing this section let us just summarize in the following table the 
diferent string contributions to the cosmological constant: 
\\
\begin{center}
\begin{tabular}{|r|l|}
\hline
$D\neq D_{Cr}$ & $\Lambda_{Cr}\sim e^{-2\Phi}\frac{D-D_{Cr}}{6\a'}$\\
Closed string divergences & $\Lambda_{c}\sim J$\\
Open string divergences &  $\Lambda_{o}\sim e^{-\Phi}\frac{1}{\a'}$\\
\hline
\end{tabular}
\end{center}


Tachyon condensation is strongly connected with these string 
contributions to the cosmological constant. Generically closed 
tachyon condensation could change the value of $\Lambda_{Cr}$ and open 
tachyon condensation, according to Sen's conjeture, can cancel 
$\Lambda_{o}$.

\section{The effective action}
\subsection{A warming up exercise}

Let us start with the following open string action
\be\label{boundaryaction}
S(a)= \int_{\Sigma}d^{2}\sigma\sqrt{h}h^{\a\b}\pd_{\a}x^{\m}\pd_{\b}x^{\n}\eta_{\m\n}+ \int_{\pd \Sigma}d\theta a
\ee

with $a$ some constant and $\Sigma$ the disc.

We will fix a world sheet metric $h_{\a\b}=e^{2\phi}\eta_{\a\b}$, thus the 
open string tachyon term in (\ref{boundaryaction}) is $ \int_{\pd \Sigma}d\theta a e^{\phi}$.

The partition function $Z(a)$ is simply defined by:
\be 
Z(a)=\int Dx e^{-S(a)}
\ee
If as usual we identify $e^{\phi}$ as the ultraviolet cutoff we get the 
beta function for $a$ :
\be\label{betafora}
\b_{a}= -a
\ee
The effective action will be defined,
 in this trivial case, by:
\be\label{zzrelation}
\frac{\pd I(a)}{\pd a}= G_{TT}\b_{a}
\ee
with $\b_{a}$ given in (\ref{betafora}) and $ G_{TT}$ the Zamolodchikov 
metric\footnote{For a formal derivation of (\ref{zzrelation}) see \cite{43}. 
Very briefly the proof is as follows. Let us define a family of two 
dimensional field theories 
\be
L=L_{0}+\lambda_{i}u^{i}(\xi)
\ee
parametrized by $\lambda_{i}$. The generating functional 
$Z(\lambda_{1}\dots \lambda_{n})$ can be expanded in powers of $\lambda$. At 
order $N$ we have
\be
Z^{N}=\int d^{2}\xi_{1}\dots d^{2}\xi_{N}<u_{n_{1}}(\xi_{1})\dots 
u_{n_{N}}(\xi_{N})>\lambda_{1}\dots \lambda_{n}
\ee
Using the OPE we get the logaritmic contribution:
\be\label{ope}
Z^{n}   =\int d^{2}\xi_{1}\dots d^{2}\xi_{N}d^{2}\xi f_{n_{1}\n_{2} m}
\frac{1}{|\xi|^{2}}\lambda_{1}\dots \lambda_{n}<u_{m}(\xi)u_{n_{3}}(\xi_{3})
\dots u_{n_{N}}(\xi_{N})>
\ee
from (\ref{ope}) we read the beta function $\b_{m}$:
\be\label{beta12}
\b_{m}=\frac{d\lambda_{m}^{R}}{d\log \Lambda}= f_{m n_{1} n_{2}}
\lambda_{n_{1}}\lambda_{n_{2}}
\ee
for $\lambda_{m}^{R}=\lambda_{m}^{B}+f_{m n_{1} n_{2}}
\lambda_{n_{1}}\lambda_{n_{2}}\log \Lambda$ with $\Lambda$ the ultraviolet 
cutoff in the integration (\ref{ope}). Defining now the effective action :
\be
\Gamma (\lambda)=\sum \lambda^{i_{1}}\dots \lambda^{i_{N}}<u_{i_{1}}
\dots u_{i_{N}}>
\ee
we get
\be\label{y_mas}
\frac{\pd \Gamma (\lambda)}{\pd \lambda_{m}}= \sum \lambda^{i_{1}}
\dots \lambda^{i_{N}}C_{e}^{i_{1}\dots i_{N}}<u_{m}u_{e}>=
\sum \beta_{e}G_{me}
\ee
where we use a generalized OPE and expression (\ref{beta12}) for 
the beta functions. In this section we will use (\ref{y_mas}) to 
define the effective action.
} defined 
by the open string amplitude on the disc of two open tachyon vertex operators 
at zero momentum: 
\be\label{gtt}
G_{TT}(a)=<1_{d},1_{d}>_{disc}(a)
\ee
with the expectation value in (\ref{gtt}) computed for the 
action (\ref{boundaryaction}).
In our case and assuming decoupling of ghosts it is obvious that $ G_{TT}$
 is equal to $e^{-a}Z(0)$. Thus using (\ref{zzrelation}) we get 
for the effective 
action $I(a)$ the following relation:
\be
\frac{\pd I(a)}{\pd a}= G_{TT}\b_{a}= -e^{-a}aZ(0)
\ee
which can be trivially integrated to:
\be
I(a)=(1+a)Z(a)=(1+a)e^{-a}Z(0)
\ee
This is a extremely interesting result since it defines a non trivial 
potential for the tachyon constant $a$ (figure 5), namely 
\be
V(a)=(1+a)e^{-a}
\ee
This potential have two extremal points at $a=\infty$ and $a=0$.
\begin{figure}[h]
\begin{center}
\leavevmode
\epsfxsize=6cm
\epsffile{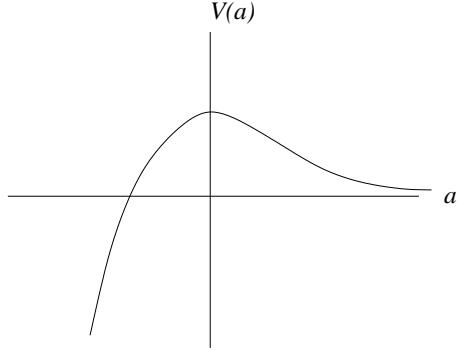}
\caption{\it Open string Tachyon potential}
\label{fig5}
\end{center}
\end{figure}


The interpretation of the two extremal points in $V(a)$ is by no means 
obvious. The extremal point $a=0$ is the standard open string vacua with 
vanishing expectation value of the open tachyon. It is a maximun 
reflecting the existence of open tachyons in the string spectrum. The 
extremal point $a=\infty$ is a bit more misterious since aparently 
it describes a stable vacua (up to tunneling processes to  $a=-\infty$)
 of the open string in flat Minkowski space time and without open tachyons.
What is the physical meaning of this vacua?

\subsection{The effective action}
Next we will consider, following ref \cite{18, 19, 20, 21, 22, 23} , a slightly more 
complicated action:
\be
S(a,u_{i})=\frac{1}{2\pi\a'}\int_{\Sigma}d^{2}\sigma\sqrt{h}h^{\a\b}
\pd_{\a}x^{\m}\pd_{\b}x^{\n}\eta_{\m\n}+ \int_{\pd \Sigma}d\theta\sqrt{h}T(x)
\ee
with
\be\label{fluctuation}
T(x)=a+\sum u_{i}x_{i}^{2}
\ee
Identifying, as usual, the ultraviolet cutoff with the world sheet 
Weyl factor we get at one loop in the $\sigma$-model
\be
\Lambda [a+u_{i}\a'\log\Lambda]
\ee
from which we derive the beta function $\b_{a}$:
\be\label{rata}
\b_{a}=-a-\sum_{i}\a' u_{i}
\ee
At this point we are interpreting $x_{i}$ in (\ref{fluctuation}) as 
representing quantum fluctuations i.e.
 $\a' u_{i}=\frac{\pd^{2}T}{\pd x_{0}\pd x_{0}}$ and $T(x_{0})=a$ 
for some background $x_{0}$. Thus we should replace in (\ref{fluctuation}) 
$u_{i}$ by $u_{i}\a'$.

In addition to $\b_{a}$ we have at tree level
\be
\Lambda [\a' u_{i}x_{i}^{2}]
\ee
which implies a beta function
\be\label{cerdo}
\b_{u_{i}}=-u_{i}
\ee
Using these tools we can define the effective action by:
\be\label{rana}
dI=\frac{\pd I}{\pd a}da+\frac{\pd I}{\pd u_{i}}du_{i}
\ee
with
\bea\label{sapo}
\frac{\pd I}{\pd a}=G_{aa}\b_{a}+G_{au_{i}}\b_{u_{i}}\\
\frac{\pd I}{\pd u_{i}}=G_{u_{i}u_{j}}\b_{u_{j}}+G_{u_{i}a}\b_{a}
\eea
where the ``metric'' factors are defined by:
\bea
G_{aa}=\int_{0}^{2\pi}d\theta<1_{d},1_{d}>_{disc} \\
G_{au_{i}}=\int_{0}^{2\pi}d\theta<1_{d},x_{i}^{2}>_{disc}\\
G_{u_{i}u_{j}}=\int_{0}^{2\pi}d\theta<x_{i}^{2},x_{j}^{2}>_{disc}
\eea
In terms of the partition function $Z(a,u_{i})$
\be
Z(a,u_{i})=\int dx e^{-S(a, u_{i})}
\ee
we get from (\ref{rana})-(\ref{sapo})
\be\label{huevo}
dI=d(\sum \a' u_{i}Z-\sum u_{j}\frac{\pd Z}{\pd u_{j}}+(1+a)Z)
\ee
where we have used:
\bea
G_{au_{i}}=\frac{\pd Z}{\pd u_{i}}\\
G_{u_{i}u_{j}}=\frac{\pd^{2}Z}{\pd u_{i}\pd u_{j}}
\eea
Integrating (\ref{huevo}) we get:
\be\label{cosa}
I=(\sum \a' u_{i}-\sum u_{j}\frac{\pd}{\pd u_{j}}+(1+a))Z(a,u_{i})
\ee
as the definition of the effective action. In this formal derivation we 
have assumed complete decoupling of ghosts. Notice that the 
contribution $1+a+\sum \a' u_{i}$ comes directly from the beta function 
$\b_{a}$ defined in (\ref{rata}) while the contribution 
$\sum u_{j}\frac{\pd}{\pd u_{j}}$ comes from the $\b_{u_{i}}$ defined 
in (\ref{cerdo}). We can rewrite (\ref{cosa}) in a more compact way 
as:
\be\label{asta}
I=(1+\b^{a}\frac{\pd}{\pd a}+\sum \b^{u_{i}}\frac{\pd}{\pd u_{i}})Z(a,u_{i})
\ee
where we have used $ Z(a,u_{i})=e^{a}\widetilde{Z}(u_{i})$.

The next step is to compute $Z(a,u_{i})$. In order to do that we need 
the Green function on the disc satisfying the boundary conditions
\be
n_{\a}\pd^{\a}x^{i}+u_{i}x^{i}=0
\ee
on $\pd\Sigma$ with $n_{\a}$ a normal vector to the boundary. This 
Green function is given by:
\be
G^{(i)}(z,w)=-\log|z-w|^2-\log|1-z\bar{w}|^{2}+\frac{2}{u}-
2u\sum_{k}\frac{1}{k(k+u)}((z\bar{w})^k+(\bar{z} w)^k)
\ee
Integrating
\be
\frac{\pd Z}{\pd u_{i}}=\int_{0}^{2\pi}d\theta<x_{i}^{2}>_{disc}
\ee
and using
\be
<x_{i}^{2}>=\lim_{\e\rightarrow 0}G_{R}^{i}(\theta, \theta +\e)
\ee
for the renormalized Green function
\be
G_{R}^{i}(\theta, \theta)=\frac{2}{u}-4u\sum_{k}\frac{1}{k(k+u)}
\ee
we get:
\be\label{truno}
Z(a,u_{i})=e^{-a}\prod_{i}\sqrt{\a' u_{i}}\Gamma(\a' u_{i})e^{\gamma\a' u_{i}}
\ee
for $\gamma$ the Euler constant. For small $u_{i}$ we can aproximate 
(\ref{truno}) by:
\be
Z(a,u_{i})\sim e^{-a}\prod_{i}\frac{1}{\sqrt{\a' u_{i}}}\qquad 
u_{i}\rightarrow 0
\ee
In this limit we get from (\ref{asta}):
\be\label{luna}
I(a,u_{i})\sim (1+a)e^{-a}\prod_{i}\frac{1}{\sqrt{\a' u_{i}}}+
\a'(\sum u_{i})e^{-a}\prod_{i}\frac{1}{\sqrt{\a' u_{i}}}+...
\ee
We can now compare the first term with:
\be
T_{25}\int d^{26}x(1+T)e^{-T}
\ee
for $T=a+\sum u_{i}x_{i}^{2}$, obtaining the well known result on the 
filling brane tension
\be
T_{25}=\frac{1}{(2\pi\a')^{13}}
\ee
Next terms in (\ref{luna}) correspond to the kinetic term for the open tachyon
\be
T_{25}\int d^{26}xe^{-T}\pd T\pd T
\ee
In order to define a potential we can change variables 
\be\label{change}
T\rightarrow \Phi=2e^{-\frac{T}{2}}
\ee
In these new variables the tachyon lagrangian becomes:
\be
S=T_{25}\int d^{26}x[\a'\pd\Phi\pd\Phi+V(\Phi)]
\ee
with (see figure 6)
\be
V(\Phi)=\frac{\Phi^2}{4}(1-\log\frac{\Phi^2}{4})
\ee
The extremal corresponding to $T=\infty$ is $\Phi=0$. The effective 
mass of the tachyon around this extremal is:
\be\label{masa}
m^2=\frac{\pd^2 V(\Phi)}{\pd\Phi\pd\Phi}|_{\Phi=0}=\infty
\ee
The extremal $T=0$ i.e. $\Phi=2$ is a maximun reproducing the standard 
open tachyon mass
\be
m^2=\frac{\pd^2 V(\Phi)}{\pd\Phi\pd\Phi}|_{\Phi=2}=-\frac{1}{\a'}
\ee
\begin{figure}[h]
\begin{center}
\leavevmode
\epsfxsize=6cm
\epsffile{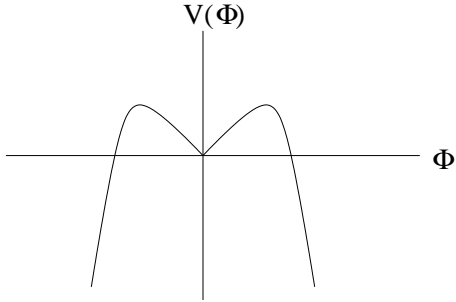}
\caption{\it Open string Tachyon potential $V(\Phi)$}
\label{fig6}
\end{center}
\end{figure}

As we see for equation (\ref{masa}) open tachyon condensation at $T=\infty$ 
induces an infinite mass for the open tachyon. Using the string mass 
formula (\ref{string_mass}) we can interpret this as an effective 
normal ordering constant $a=-\infty$. If we do that the dominant 
contribution to the BRST charge (\ref{brst_charge}) is just the 
comohologicaly trivial BRST charge $Q=c_{0}$.

This heuristic argument indicates in agreement with Sen's conjeture that no 
open string degrees of freedom survive once the tachyon condenses to 
$T=\infty$. In summary we can interpret the vacuum defined by the 
$T=\infty$ condensate as the closed string vacua. The closed string 
tachyon can be interpreted as associated with the quantum unstability due to 
tunneling processes from $\Phi=0$ to $\Phi=\infty$

\subsection{Non-critical dimension and tachyon condensation}
The space time lagrangian for the open tachyon is given by:
\be
S=T_{25}\int d^{26}xe^{-T}[\a'\pd T\pd T+(1+T)]
\ee
The corresponding equation of motion is
\be\label{ecuacion}
2\a'\pd^{\m}\pd_{\m}T-\a'\pd^{\m}T\pd_{\m}T+T=0
\ee
A soliton solution for the equation (\ref{ecuacion}) is given by:
\be
T(x)=a+\sum u_{i}x_{i}^{2}
\ee
with $u_{i}=\frac{1}{4\a'}$ or $u_{i}=0$ and $a=-n$ for $n$ the number 
of non vanishing $u_{i}$'s. In terms of the field $\Phi$ defined 
in (\ref{change}) the profile of the solution looks like the one 
depicted in figure 7.
\begin{figure}[h]
\begin{center}
\leavevmode
\epsfxsize=7cm
\epsffile{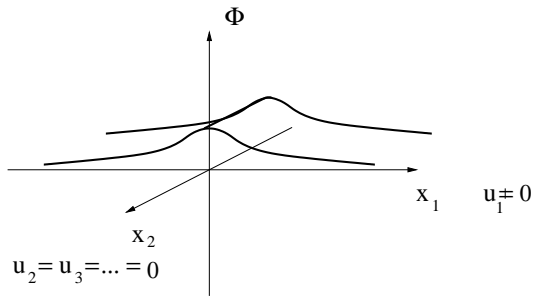}
\caption{\it Soliton shape}
\label{fig7}
\end{center}
\end{figure}

This can be interpreted in a first approximation as $D-(25-n)$ soliton
brane.

In principle we can try to play the same game but including the effect 
of a non trival dilaton. The simplest example will be of course 
to work with non critical dimension $n$ and a linear dilaton background
\be
\Phi=qy
\ee
with $q=\sqrt{\frac{n-26}{6\a'}}$. Inspired by the Liouville picture 
of non critical strings we take the linear dilaton depending only on one 
coordinate $y$. The lagrangian including the effect of the dilaton would 
be most like:
\be
S=T_{25}\int d^{26}xe^{-\Phi}e^{-T}[\a'\pd T\pd T+(1+T)]
\ee
The equation of motion becomes:
\be
2\a'\pd^{\m}\pd_{\m}T-\a'\pd^{\m}T\pd_{\m}T-\a'\pd^{\m}\Phi\pd_{\m}T+T=0
\ee
As solution we can try 
\be
T(x)=a+\sum u_{i}x_{i}^{2}\qquad a=-m\qquad u_{i}=\frac{1}{4\a'}\qquad i=1...m
\ee
with $u_{y}=0$. This soliton defines a $D-(n-m-1)$-brane that extends on the 
``Liouville''-direction. Notice that we have not soliton solutions for 
$u_{y}\neq 0$ which seem to imply that tachyon condensation is not 
taking place in the Liouville direction. This lead us to suggest 
the following conjecture: In non critical open strings open tachyon 
condensation can not take place in the Liouville direction.


A trivial corolary of the previous conjecture is that in space-time 
dimension equal two tachyon condensation does not take place, 
which is consistent with the fact that tachyons in $D=2$ with the 
linear dilaton turn on are not real tachyons.

\section{D-branes, tachyon condensation and K-theory}

\subsection{Extended objects and topological stability}

Let us start considering a gauge theory with a Higgs field $\Phi$:
\be\label{Higgs_model}
L=L_{0}(A^{\m},\Phi)+V(\Phi)
\ee
for some Higgs potential $V(\Phi)$. Necessary conditions for 
the existence of topologically stable extended objects of space codimension 
$p$ is the non triviality of the homotopy group
\be\label{homotopy_group}
\Pi_{p-1}(V)
\ee
for $V$ the manifold of classical vacua of lagrangian (\ref{Higgs_model}).

In fact for an extended object of codimension $p$ the condition of finite 
density of energy implies that at the infinity region in the transversal 
directions -whose topology is $S^{p-1}$- the field configuration must 
belong to the vacuum manifold $V$. Hence we associate with each 
configuration of finite density of energy a map
\be
\Psi: S^{p-1}\rightarrow V
\ee 
whose topological clasification is defined by the homotopy group 
(\ref{homotopy_group}).

The simplest example of vacuum manifold corresponding to spontaneous 
breaking of symmetry $G\rightarrow H$ is the homogeneous space
\be 
V=G/H
\ee

So 'tHooft-Polyakov monopole for instance is defined for $G=SU(2)$ and 
$H=U(1)$ by the topological condition $\Pi_{2}(G/H)=Z$ which 
coincide with its magnetic charge.

\subsection{A gauge theory analog for D-branes in type II strings}

We know that in type II strings we have extended objects which are RR 
charged and stable, namely the D-branes. For type IIA we have 
$D_{p}$-branes with $p$ even and for type IIB $D_{p}$-branes with $p$ odd. 
Since we are working in critical 10 dimensional space-time the space 
codimension of those $D_{p}$-branes is odd $2k+1$ for type IIA and even $2k$ 
for type IIB.

We will consider now the following formal problem. Try to get two 
gauge Higgs lagrangians $L^{IIA(IIB)}(A_{\m},\Phi)$ such that it 
can be stabilished a one to one map between type II D-branes and 
topological stable extended objects for those lagrangians in the 
sense defined in previous section. We will denote this formal gauge 
theory the \underline{gauge theory analog} of the type II strings.

Of course the hint for answering this question is Sen's tachyon 
condensation conjecture for type II strings. We will present 
first this construction in the case of type IIB strings.

\subsubsection{Sen's conjecture for type IIB}

In type IIB strings we have well defined $D_{9}$ filling branes. Since 
they are charged under the RR sector we can define the corresponding 
$D_{\bar{9}}$-antibranes. As it is well known the low energy physics on 
the world volume of a set of $N$  $D_{9}$-branes is a $U(N)$ gauge theory 
without open tachyons. In fact the open tachyon is projected out by the 
standard GSO projection
\be
(-1)^{F}=+1
\ee
for $F$ the world sheet fermion number operator. The situation changed 
when we consider $N$  $D_{9}$-branes and $N$ $D_{\bar{9}}$-antibranes. 
In this case the theory on the world volume is $U(N)\times U(N)$ and not 
$U(2N)$ due to the fact that the GSO projection on open string states with 
end points at a $D_{9}$-brane and a $D_{\bar{9}}$-antibrane is the opposite, 
namely
\be\label{GSO}
(-1)^{F}=-1
\ee

This projection eliminates from the spectrum the massless gauge vector 
bosons that will enhance the $U(N)\times U(N)$ gauge symmetry to $U(2N)$. 
In addition the projection (\ref{GSO}) is not killing the tachyon in the 
$(9,\bar{9})$ and $(\bar{9},9)$ open string sectors. Thus the 
gauge theory associated with the configuration of 
$N$  $D_{9}$-branes and $N$ $D_{\bar{9}}$-antibranes is a $U(N)\times U(N)$ 
gauge theory with a Higgs field, the open tachyon, in the bifundamental 
$(N,\bar{N})$ representation.

This gauge theory will be a natural starting point for defining the gauge 
analog  model in the case of type IIB strings.

Of course in order to get a rigurous criterion for the topological 
stability of extended objects in this gauge theory we need to 
know the potential for the open tachyon. This potential is something that at 
this point we don't know how to calculate in a rigorous way. However we 
can assume that a tachyon condensation is generated with a vacuum expectation 
value
\be
<T>=T_{0}
\ee
with $T_{0}$ diagonal and with equal eigenvalues. If this condensate 
takes place then the vacuum manifold is simply
\be
V=\frac{U(N)\times U(N)}{ U_{D}(N)}\sim  U(N) 
\ee

Thus the condition for topological stability for extended objects of space 
codimension $2k$ will be:
\be
\Pi_{2k-1}(U(N))\neq 0
\ee
which by Bott periodicity theorem:
\be\label{Bott}
\Pi_{j}(U(k))=\left\{ \begin{array}{ll}
Z & \textrm{j odd $j<2k$}\\
0 & \textrm{j even $j<2k$}
\end{array} \right.
\ee
we know is the case for big enough $N$.

The simplest example will be to take $k=1$ corresponding to the extended 
object of the type of a $D_{7}$-brane. The condition of finite energy 
density defines a map from $S^{1}$ into $U(N)$. For just one pair of 
$D_{9}-D_{\bar{9}}$ configuration we get $\Pi_{1}(U(1))=Z$, with this winding 
number representing the ``magnetic charge'' of the  $D_{7}$-brane that looks 
topologically like a vortex line (see figure 8).


\begin{figure}[h]
\begin{center}
\leavevmode
\epsfxsize=7cm
\epsffile{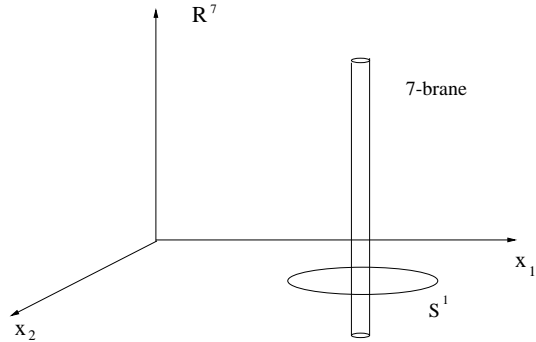}
\caption{\it Topology of the $D_{7}$-brane }
\label{fig8}
\end{center}
\end{figure}


If we go to the following brane namely the $D_{5}$-brane we have $k=2$ 
and we need a non vanishing homotopy group $\Pi_{3}(U(N))$. The 
minimun $N$ for which this is possible according to (\ref{Bott}) is $N=2$ 
i.e. two pairs of $D_{9}-D_{\bar{9}}$-branes. We can understand what is 
happenig in two steps. First of all we get a configuration of two 
$D_{7}$-branes and from this the $D_{5}$-brane.

For $k=3$ we need non vanishing homotopy group $\Pi_{5}(U(N))$. 
The natural $N$ we should choose is dictated by the step construction 
namely $N=4$. In general for codimension $2k$ we will consider a gauge group 
$U(2^{k-1})$.

\subsection{K-theory version of Sen's conjecture}

The configuration of $D_{q}-D_{\bar{q}}$-branes naturally defines a couple 
of $U(N)$ vector bundles $(E,F)$ on space time. The main idea of Sen's 
tachyon condensation is that a configuration characterized by the couple 
of vector bundles $(E,E)$ with a topologically trivial tachyon 
field configuration decays into the closed string vacua for type IIB string 
theory.This is exactly the same type of  phenomena we have discussed 
in section 7 for the bosonic string.  The phenomena naturally leads 
to consider, as far as we are concerned with $D$-brane charges, instead 
of the couple of bundles  $(E,F)$ the equivalence class defined by \cite{31}
\be\label{equivalence_relation}
(E,F)\sim (E\oplus G,F\oplus G)
\ee  
for $\oplus$ the direct sum of bundles. This is precisely the definition 
of the K-group of vector bundles on the space-time $X$, $K(X)$.

Let us here recall that the space $A=V_{ec}(X)$ of vector bundles on $X$ 
is a semigroup with respect to the operation of direct sum. The way 
to associate with $A$ a group $K(A)$ is as the quotient space in
$A\times A$ defined by the equivalence relation
\be
(m,n)\sim (m',n') 
\ee
if $\exists p$ such that $m+n'+p=n+m'+p$ which is precisely what we are doing 
in the definition (\ref{equivalence_relation}).

A different but equivalent way to define $K(A)$ for $A=V_{ec}(X)$ is 
as the set of equivalence classes in $V_{ec}(X)$ defined by the 
equivalence relation:
\be
E\sim F \textrm{ if } \exists G: E\oplus G= F\oplus G 
\ee
where the ``$=$'' means isomorphism.

It is convenient for our purposes to work with the reduced K-group 
$\tilde{K}(X)$ which is defined by
\be
Ker[K(X)\rightarrow K(p)]
\ee
for $p$ a point in $X$. Notice that $K(p)$ is just the group of integer 
numbers $Z$. This is the group naturally associated with the semigroup 
$V_{ec}(p)=N$ where $N$ here parametrizes the different dimensions of the vector bundles in $V_{ec}(p)$.

In order to characterize $D_{p}$-branes in type IIB in terms of K-theory 
we will need to consider the group $K(X,Y)$. We will consider $X$ a 
compact space and $Y$ also compact and contained in $X$.

In order to define $K(X,Y)$ we will use triplets $(E,F,\a)$ where $E$ 
and $F$ are vector bundles on $X$ and where $\a$ is an isomorphism:
\be
\a : E\mid_{Y}\rightarrow F\mid_{Y}
\ee
of the vector bundles $E$ and $F$ reduced to the subspace $Y$ \cite{36}.

The definition of $K(X,Y)$ requieres to define 
\underline{elementary triplets}. An elementary triplet is given by 
$(E,F,\a)$ with $E=F$ and $\a$ homotopic to the identity in the space 
of automorphisms of $E\mid_{Y}$. Once we have defined elementary 
triplets the equivalence relation defining $K(X,Y)$ is :
\be
\sigma = (E,F,\a) \qquad  \sigma' =(E',F',\a')
\ee
$\sigma \sim \sigma'$ iff $\exists $ elementary triplets $\tau$ and $\tau'$ 
such that 
\be
\sigma + \tau = \sigma' + \tau'
\ee
where
\be
\sigma + \tau =  (E\oplus G,F\oplus G, \a\oplus 1_{d})
\ee

Once we have defined $K(X,Y)$ we can try to put in this language the 
topological characterization of a $D_{p}$-brane of space 
codimension  $2k$. Namely we will take as $Y$ the ``boundary'' region 
in transversal directions i.e. $S^{2k-1}$. As space $X$ we will take the ball 
$B^{2k}$. The tachyon field transforming in the bifundamental 
representation will define on $S^{2k-1}$ an isomorphism between the 
two vector bundles $E,F$ defined by the starting configurations of 
$D_{9}-D_{\bar{9}}$-branes. Finally the homotopy class of this map will 
define the charge of the  $D_{q}$-brane 
of space codimension $2k$. The K-group we define in this way is
\be
K(B^{2k},S^{2k-1})
\ee

Now we can use the well known relation:
\be
K(B^{2k},S^{2k-1})=\tilde{K}(B^{2k}/S^{2k-1})
\ee
where $X/Y$ is defined by contracting $Y$ to a point. 

It is easy to see that 
\be
B^{2k}/S^{2k-1}\sim S^{2k}
\ee

Thus we can associate with type IIB $D_{p}$-branes of space codimension $2k$ 
elements in $\tilde{K}(S^{2k})$.

In order to define the tachyon field in this case we will 
specify the isomorphism $\a$. For codimension $2k$ let us consider the 
$2^{k}\times 2^{k}$ gamma matrices $\Gamma_{i}\quad (i=0\dots 2k)$. Let $v$ 
a vector in $C^{2k}$. The isomorphism $\a$ is defined by
\be
\a (x,v)= (x,x_{i}\Gamma^{i}(v))
\ee
for $x\in S^{2k-1}$. The tachyon field is defined by
\be
T(x)\mid_{x\in S^{2k-1}}= x_{i}\Gamma^{i}
\ee

\subsection{Type IIA strings}

Next we will define a gauge analog for type IIA strings. The gauge-Higgs 
lagrangian will be defined in terms of a configuration of $D_{9}$-branes.
For type IIA $D_{9}$-branes are not BPS and therefore they are unstable. 
The manifestation of this unstability is the existence of an open tachyon 
field $T$ transforming in the adjoint representation. The gauge group 
for a configuration  of $N$  $D_{9}$-branes is   $U(N)$. Notice that in 
the case of the type IIA we can not use $D_{\bar{9}}$-antibranes since 
for type IIA $D_{9}$-branes are not RR-charged.

We can now follow the same steps that in the type IIB case, namely to 
look for a tachyon potential and to compute the even homotopy groups of 
the corresponding vacuum manifold. Instead of doing that we will approach 
the problem from a different point of view, interpreting the D-branes of 
type IIA as topological defects associated with the gauge fixing topology. 
In order to describe this approach we need first to review some know facts 
about gauge fixing topology for non abelian gauge theories.

\subsubsection{'tHooft's abelian projection}\label{abelian_proyection}
An important issue in the quantization of non abelian gauge theories 
is to fix the gauge. By an unitary gauge we means a procedure to 
parametrize the space of gauge ``orbits'' i.e. the space of physical 
configurations 
\be
R/G
\ee
for $R$ the total space of field configurations, in terms of physical 
degrees of freedom where by that we mean those that contribute to the 
unitary S-matrix. This in particular means a ghost free gauge fixing.

In reference \cite{38} 'tHooft suggested a way to fix the non 
abelian gauge invariance in a unitary way. This type of gauge fixing 
known as ``abelian projection'' reduce the physical degrees of freedom to 
a set of $U(1)$ photons and electrically charged vector bosons.

In addition to these particles there is an extra set of dynamical 
degrees of freedom we need to include in order to have a complete 
description of the non abelian gauge theory. These extra degrees of 
freedom are magnetic monopoles that appear as a consequence of the 
topology of the gauge fixing.

More precisely let $X$ be a field transforming in the adjoint representation
\be
X\rightarrow gXg^{-1}
\ee
The field $X$ can be a functional $X(A)$ of the gauge field $A$ or 
some extra field in the theory. The way to fix the gauge is to impose $X$ 
to be diagonal
\be\label{1}
X=
\left(
\begin{array}{ccc}
\lambda_{1}&~&~\\
~&\ddots&~\\
~&~&\lambda_{N}\\
\end{array}
\right)
\ee

The residual gauge invariance for a $U(N)$ gauge theory is $U(1)^{N}$ i.e. 
gauge transformations of the type
\be\label{2}
g=
\left(
\begin{array}{ccc}
e^{i\a_{1}}&~&~\\
~&\ddots&~\\
~&~&e^{i\a_{N}}\\
\end{array}
\right)
\ee

The degrees of freedon of this gauge are:
\begin{itemize}
\item $N$-$U(1)$ photons
\item $\frac{1}{2}N(N-1)$ charged vector bosons 
\item $N$ scalars fields $\lambda_{i}$
\end{itemize}

Gauge fixing singularities will appear whenever two eigenvalues coincide
\be
\lambda_{i}=\lambda_{i+1}
\ee
Notice that we can fix the gauge impossing $X$ to be diagonal and that 
$\lambda_{i}>\lambda_{i+1}>\lambda_{i+2}>\dots$.
What is the physical meaning of these gauge fixing singularities?

First of all it is easy to see that generically these gauge fixing 
singularities have codimension 3 in space. In particular this 
means that if we are working in four dimensional space time they 
behave as pointlike particles.

Secondly if we consider the field $X$ in a close neighborhood of the 
singular point before gauge fixing 
\be\label{3}
X=
\left(
\begin{array}{c|c|c}
D_{1}&0&0\\\hline
0&
\begin{tabular}{cc}
$\lambda + \epsilon_{3}$&$\epsilon_{1}-i \epsilon_{2}$\\
$\epsilon_{1}+i \epsilon_{2}$&$\lambda-\epsilon_{3}$\\
\end{tabular}
&0\\\hline
0&0&D_{2}\\
\end{array}
\right)
\ee
we can write the small two by two matrix in (\ref{3}) as:
\be
X=\lambda 1_{d}+\e_{i}\sigma_{i}
\ee
for $\sigma_{i}$ the Pauli matrices. The field $\e (x)$ is equal to zero 
at the singular point and in a close neighborhood can be written as:
\be
\e (x)=\sum_{i=1}^{3}x_{i}\sigma_{i}
\ee

We can easily relate this field to a magnetic monopole. In fact let us 
consider $S^{2}$ in $R^{3}$ and let us define the field on $S^{2}$:
\be\label{extra_field}
X(x)\mid_{x\in  S^{2}}= \sum_{i=1}^{3}x_{i}\sigma_{i}
\ee
Clearly $X^{2}(x)\mid_{x\in  S^{2}}=1_{d}$, thus we can define the projector:
\be
\Pi_{\pm}=\frac{1}{2}(1 \pm X(x))
\ee
The trivial bundle $S^{2}\times C^{2}$ decomposse into:
\be
S^{2}\times C^{2}=E_{+}\oplus E_{-}
\ee
where the line bundles $E_{\pm}$ are defined by the action of the projection 
$\Pi_{\pm}$ on $C^{2}$. The associated principal bundle to $E_{+},E_{-}$ 
define the magnetic monopoles.

In summary the gauge fixing singularities of gauge (\ref{1}) corresponding 
to two 
equal eigenvalues should be interpreted as pointlike magnetically charged 
particles. It is important to stress that the existence of these magnetic 
monopoles is completly independent of being in a Higgs or confinement phase.

\subsubsection{The $D_{6}$-brane}
Here we will repeat the discussion in \ref{abelian_proyection} but 
for the $U(N)$ gauge theory defined by a configuration of  $D_{9}$ unstable 
filling branes. We will use the open tachyon field transforming 
in the adjoint representation to fix the gauge. By impossing $T$ 
to be diagonal we reduce the theory to pure abelian degrees of freedom 
in addition to magnetically charged objects of space codimension 3 
that very likely can be identified with $D_{6}$-branes.

Using expresion (\ref{extra_field}) and replacing the $X$ field by the 
open tachyon we find that in the close neighborhood of a codimension 3 
singular region the tachyon field is represented by:
\be
T(x)\mid_{x\in  S^{2}}= \sum_{i=1}^{3}x_{i}\sigma_{i}
\ee
which is precisely the representation of the tachyon field around a 
$D_{6}$-brane suggested in reference \cite{32}.

\subsubsection{K-theory description}

The data we can naturally associate with a configuration of $D_{9}$-branes 
in type IIA is a couple $(E,T)$ with $E$ a vector bundle and $T$ the open 
tachyon field. We will translate these data into a more mathematical 
language using the higher K-group $K^{-1}(X)$ \cite{31, 32}.

In order to define $K^{-1}(X)$ we will start with couple $(E,\a)$ with 
$E$ a vector bundle on $X$ and $\a$ an automorphism of $E$. As we did 
in the definition of  $K(X,Y)$ we define elementary pairs $(E,\a)$ if 
$\a$ is homotopic to the identity within automorphisms of $E$. Using 
elementary pairs $(E,\a)=\tau$ we define the equivalence relation 
\be
\sigma \sim \sigma'
\ee
iff $\exists \tau ,\tau'$ elementary such that
\be
\sigma \oplus \tau = \sigma' \oplus \tau'
\ee
We can now define $K^{-1}(X,Y)$ as pairs $(E,\a)\in K^{-1}(X) $ such that 
$\a \mid_{Y}=1_{d}$.

As before we will use the tachyon field $T$ to define the automorphism $\a$. 
In codimension 3 we got in the previous section that 
\be
T(x)\mid_{x\in  S^{2}}= \sum_{i=1}^{3}x_{i}\sigma_{i}
\ee
Clearly $T^{2}(x)\mid_{x\in  S^{2}}=1_{d}$ and therefore if we define
\be
\a= e^{iT}
\ee
and we identify $Y=S^{2}$ we get the condition 
\be
\a \mid_{Y}=1_{d}
\ee
used in the definition of  $K^{-1}(X,Y)$. Thus we associate the 
$D_{p}$-branes of codimension $2k+1$ with elements in    
\be
K^{-1}(B^{2k+1},S^{2k})
\ee

Using again the relation 
\be
K^{-1}(B^{2k+1},S^{2k})=\tilde{K}^{-1}(B^{2k+1}/S^{2k})
\ee
and
\be
\tilde{K}^{-1}(X)=\tilde{K}(SX)
\ee
for $SX$ the reduced suspension of $X$ (in particular $SS^{n}=S^{n+1}$) 
we conclude that $D_{p}$-branes in type IIA are associated with
\be
\tilde{K}^{-1}(S^{2k+1})= \tilde{K}(S^{2k+2})
\ee

The reader can wonder at this point in what sense to work with K-theory 
is relevant for this analysis. The simplest answer comes from remembering the 
group structure of $K^{-1}(X)$.

The group structure of in $K^{-1}(X)$ is associated with the definition 
of inverse. Namely the inverse of $(E,\a)$ is $(E,\a^{-1})$. The reason is 
that
\be
(E,\a)\oplus (E,\a^{-1})=(E\oplus E, \a \oplus \a^{-1})
\ee
where
\be\label{4}
\a \oplus \a^{-1}=
\left(
\begin{array}{cc}
\a&0\\
0&\a^{-1}\\
\end{array}
\right)
\ee

Now there is a homotopy transforming matrix (\ref{4}) into the identity           

\be\label{5}
\left(
\begin{array}{cc}
\a&0\\
0&\a^{-1}\\
\end{array}
\right)(t)=
\left(
\begin{array}{cc}
\a&0\\
0&1\\
\end{array}
\right)
\left(
\begin{array}{cc}
\cos{t}&\sin{t}\\
-\sin{t}&\cos{t}\\
\end{array}
\right)
\left(
\begin{array}{cc}
1&0\\
0&\a^{-1}\\
\end{array}
\right)
\left(
\begin{array}{cc}
\cos{t}&\sin{t}\\
\sin{t}&\cos{t}\\
\end{array}
\right)
\ee
such that 
\be\label{6}
\left(
\begin{array}{cc}
\a&0\\
0&\a^{-1}\\
\end{array}
\right)(t=1)=
\left(
\begin{array}{cc}
\a\a^{-1}&0\\
0&1\\
\end{array}
\right)= 1_{d}
\ee

What this homotopy means is again Sen's tachyon condensation conjecture. In 
fact if we associate a $D_{6}- D_{\bar{6}}$ brane configuration with a 
matrix $T$ with two pairs of eigenvalues  $(\lambda_{i}=\lambda_{i+1})$ and 
$(\lambda_{j}=\lambda_{j+1})$ equal. This configuration is - because of 
homotopy (\ref{5}) - topologically equivalent to the vacuum. We see once more 
how Sen's condensation is at the core of the K-theory promotion of 
$V_{ec}(X)$ from a semigroup into a group.

\section{Some final comments on gauge theories}
The data associated with a gauge theory and a unitary gauge fixing of the 
type used in the abelian projection can be summarized in the same type 
of couples used to define the higher K-group $K^{-1}(X)$, namely a $U(N)$ 
vector bundle $E$ and an automorphism $\a$. In this sense gauge fixing 
topology is traslated into the homotopy class of $\a$ within the 
automorphisms of $E$. In standard four dimensional gauge theories 
the gauge fixing topology is described in terms of magnetic monopoles and 
antimonopoles. In principle we have different types of magnetic monopoles 
charged with respect to the different $U(1)$'s in the Cartan subalgebra. 
The group theory meaning of $K^{-1}(X)$ is reproduced, at the gauge theory 
level, by the homotopy (\ref{5}) that is telling us that 
monopole-antymonopole pairs, although charged with respect
to different $U(1)$'s in the Cartan subalgebra, annihilate 
into the vacuum, very much in the same way as, by Sen's tachyon 
condensation, a pair of brane-antibrane decay into the vacuum.

In what we have denoted the gauge theory analog of type II strings, 
namely a gauge-Higgs lagrangian with topologically stable extended objects 
in one to one correspondence with type II stable 
$D_{p}$-branes, it is apparently absent one important dynamical aspect 
of $D$-filling brane configurations. In fact in the case of unstable 
filling branes the decay into the vacuum comes together with the 
process of cancelation of the filling brane tension and thus with the 
``confinement'' of ``electric'' open string degrees of freedom. The 
resulting state is a closed string vacua with stable $D_{p}$-branes that 
are sources of RR fields which are part of the the closed string spectrum. 
The dynamics we lack in the gauge theory analog is on one side the 
equivalent of the confinement of open string degrees or freedom
\footnote{
i.e. the concept of a filling brane tension which is intimately related 
to the open-closed relation in string theory, namely to soft dilaton 
absortion to the vacuum} and on the other side the RR closed string 
interpretation of the dual field created by the $D_{p}$-brane 
topological defects. Very likely the gauge theory interpretation of this two 
phenomena can shed some light on the quark confinement problem.

\section*{Acknowledgments}
C.G. thanks the organizers of the IV SIGRAV School at Como for giving me the 
opportunity to present these lectures. We also thank E. Alvarez for useful 
discussions. This work is supported by the Spanish grant AEN2000-1584. The 
work of P.R. is supported by a FPU U.A.M. grant.


\end{document}